\documentstyle [11pt] {article}
\begin{document}
\baselineskip 16pt

\centerline{\Large \bf More General Quantum Search Algorithm $Q=-I_{\gamma }VI_{\tau }U$ } 
 \centerline{\Large\bf And the Precise Formula for the Amplitude  and} 
\centerline{\Large\bf the Non-symmetric Effects of Different Rotating Angles }
\footnote{ The paper was supported by NSFC and partially by Lab. of computer science, ISCAS }

\centerline { Dafa Li}

\centerline{ Dept. of Mathematical Sciences}
\centerline{ National Lab. for AI at Tsinghua University}
\centerline{ Tsinghua University Beijing 100084 CHINA}
\centerline{email: dli@math.tsinghua.edu.cn}
\centerline{ Xinxin Li}

\centerline{ Dept. of Computer Science}
\centerline{ Wayne State University, Detroit, MI 48202, USA}
\bigskip

\begin{abstract}

 This paper presented two  general quantum search algorithms. 
We derived the iterated formulas and the simpler approximate formulas  
and the precise formula for the amplitude in the desired state.
A mathematical proof of Grover's algorithm being
 optimal among the algorithms with arbitrary phase
rotations was given in this paper. 
This first reported the non-symmetric effects of different rotating angles, 
and gave the first-order approximate phase condition when rotating angles are
different.
\end{abstract}



\section{Introduction}

\bigskip

Shor reported his prime factoring algorithm in \cite{Shor}. Then Grover
gave his quantum search algorithm \cite{Grover971}\cite{Grover972}.
In \cite{Grover971} Grover's quantum search algorithm consists of a
sequence of unitary operations on a pure state. The algorithm is $%
Q=-I_{0}^{(\pi )}WI_{\tau }^{(\pi )}W,$where $I_{x}^{(\pi
)}=I-2|x\left\rangle |\right\langle x|$ and  inverts
the amplitude in the state $|x\rangle $,  $W$ is Walsh-Hadamard
transformation. It would carry out repeated operations of $Q,$that is, $%
...(-I_{0}^{(\pi )}WI_{\tau }^{(\pi )}W)(-I_{0}^{(\pi )}WI_{\tau }^{(\pi
)}W)(-I_{0}^{(\pi )}WI_{\tau }^{(\pi )}W)...$

$=...(-WI_{0}^{(\pi )}W)I_{\tau }^{(\pi )}(-WI_{0}^{(\pi )}W)I_{\tau }^{(\pi
)}(-WI_{0}^{(\pi )}W)I_{\tau }^{(\pi )}...,$
where  $(-WI_{0}^{(\pi )}W)$ is just
the inversion-about-average operation. Therefore Grover's algorithm consists
of alternating iteration of $(-WI_{0}^{(\pi )}W)$, inversion about the
average, and $I_{\tau }^{(\pi )},$inversion\ of the amplitude in the desired
state $|\tau \rangle $. Then in \cite{Grover98} Grover showed how to
replace Walsh-Hadamard transformation in his original algorithm with an arbitrary
quantum mechanical operation,  obtained the quantum search algorithm $%
Q=-I_{\gamma }^{(\pi )}U^{-1}I_{\tau }^{(\pi )}U,$where $U$ is an arbitrary unitary
operation and $U^{-1}$ is the adjoint (the complex conjugate of the
transpose) of $U$. Grover thinks that it leads to several new applications and
broadens the scope for implementation. When the amplitudes are rotated by
arbitrary phases, instead of being inverted, that is, $Q=-I_{\gamma
}^{(\theta )}U^{-1}I_{\tau }^{(\phi )}U$ ,where $I_{x}^{(\theta
)}=I-(-e^{i\theta }+1)|x\left\rangle |\right\langle x|,$ Long  et al. %
 first found \ $\theta$ and $\phi$ must satisfy a matching condition: $\theta =\phi$,
and  derived an approximate formula for the amplitude 
in the desired state\cite{Long1} and studied the effects of using arbitrary
phases in amplitude amplification \cite{Long3}. In
\cite{Hoyer} the phase condition that $\tan (\varphi /2)=\tan (\phi /2)(1-a)$ was
presented. In \cite{Bilham} the recursion equations were used to study the quantum
search algorithm. 

Here, this paper presented two  general quantum
search algorithms, derived the iterated formulas and the simpler
approximate formulas of the amplitudes in the desired state. 
We showed that the
amplitude in the desired state 
can be precisely  written as a polynomial
in  $(\beta \lambda)$
for any quantum search algorithms   which
preserve  a two-dimensional vector space.  
This is the first precise formula
of the amplitude amplification in the desired state 
for the general quantum search algorithms
 including Grover' and Long et al.' and Hoyer's ones.
The precise formula for the amplitude
can  help derive a precise phase condition.
A mathematical proof was given in this paper for that  Grover's algorithm is
optimal among the quantum search algorithms with arbitrary phase
rotations. We first found that the effects of rotating angles in the initial
state  and the desired state  on the
amplitude of the desired state are not symmetric. 
We also discussed the amplitude
amplifications, and gave the first-order approximate phase condition when
rotating angles are different. 

\section{
More general quantum search algorithm $Q=-I_{\gamma }VI_{\tau }U$ }

Let $I_{x}^{(\theta )}=I-(-e^{i\theta}+1)|x\left\rangle |\right\langle x|$. 
Grover studied the quantum search algorithm in [3]: $Q=-I_{\gamma }^{(\pi
)}U^{-1}I_{\tau }^{(\pi )}U,$
 where $U$ is an arbitrary unitary operator and $U^{-1}$ is
 the adjoint (the complex conjugate of the transpose) of $U$ and $%
I_{x}^{(\pi )}$ inverts the amplitude in the state $|x\rangle $. 
 Generally let 
$I_{x}=I-ae^{i\theta }|x\left\rangle |\right\langle x|$. Then $I_{x}$ is
unitary if and only if $(1-ae^{i\theta })(1-ae^{-i\theta })=1.$ That is, $a=2\cos
\theta $. Then $I_{x}=I-2\cos \theta e^{i\theta }|x\left\rangle
|\right\langle x|.$When $\theta =0$, $I_{x}=I_{x}^{(\pi )}$. If let $%
I_{x}^{\prime }=I-(ae^{i\theta }+1)|x\left\rangle |\right\langle x|,$ then $%
I_{x}^{\prime }$ is unitary if and only if $a=\pm 1.$That is, $I_{x}^{^{\prime
}}=I_{x}^{(\theta )}$ or $I_{x}^{(\pi +\theta)}$.
 Clearly $%
I_{x}^{^{(\pi +\theta )}}=I_{x}^{(\pi )}I_{x}^{(\theta )}$ and $I_{x}^{(\pi
+\theta )}=I-2\cos \frac{\theta }{2}e^{i\theta }|x\left\rangle
|\right\langle x|$ and $I_{x}=$ $I-2\cos \theta e^{i\theta }|x\left\rangle
|\right\langle x|=I_{x}^{(\pi +\theta )}I_{x}^{(\theta )}=I_{x}^{(\pi
)}(I_{x}^{(\theta )})^{2}.$ Long et al. studied the phase matching
condition for the algorithm \cite{Long1} $Q=-I_{\gamma }^{(\theta
)}U^{-1}I_{\tau }^{(\phi )}U$.

Let's study the quantum search algorithm $Q=-I_{\gamma }VI_{\tau }U,$ where $%
V $ and $U$ are arbitrary unitary  $N\times N$ matrices , 
where $N=2^{n}$ for $n$ qubits, and  $I_{\gamma }=I-2\cos \theta e^{i\theta
}|\gamma \left\rangle |\right\langle \gamma |$ and $I_{\tau }=I-2\cos \phi
e^{i\phi }|\tau \left\rangle |\right\langle \tau |.$

Since $V$ and $U$ are unitary, $(VU)$ is also unitary,
where $(VU)$ is the product of $V$ by $U$.  
 Then by the definition $(VU)^{-1}=(VU)^{+}$ \cite{Birkhoff}
 which is the adjoint (the complex conjugate of the transpose) of $(VU)$. 
Assume that $(VU)$ is hermitian. By the definition $(VU)=$ $(VU)^{+}$ \cite{Birkhoff},
 therefore $(VU)(VU)=$ $(VUVU)=I,$
where $(VUVU)$ is the product of the matrices.
In fact, given that $(VU)$ is unitary 
then $(VU)$ is hermitian if and only if $(VUVU)=I.$
We give a proof as follows. Since $(VU)$ is unitary and hermitian,
$(VU)(VU)=(VU)(VU)^+=I$. Conversely, from that $(VUVU)=I,$ we obtain 
$(VU)(VU)=(VU)(VU)^+$ since $VU$ is unitary,
then $(VU)=(VU)^+$. Therefore by the definition  $VU$ is hermitian.

For Grover's algorithm $V=U^{-1}$
clearly $(VU)$  is hermitian.
Here we give a sufficient condition in which $(VU)$ is hermitian as follows.
If $V$ and $U$ are hermitian and $V$ and $U$ are commutative, 
then $(VU)$ is hermitian since by the definition $V=V^+$ and 
$U=U^+$ and $(VU)^{+}=$ $U^+V^+=UV=VU.$

On the assumption that $(VU)$ is hermitian  we will show
 that $Q$ preserves the four - dimensional
vector space spanned by $|\gamma \rangle $, $(VU)|\gamma \rangle $, $V|\tau
\rangle $ and $U^{-1}|\tau \rangle .$ After  obtain $V|\tau \rangle $ and $%
U^{-1}|\tau \rangle ,$ we apply $V^{-1}$ to $V|\tau \rangle $ or $U$ to $%
U^{-1}|\tau \rangle,$ then obtain the desired state $|\tau \rangle $. Let's
calculate the amplitude in the desired state $V|\tau \rangle $ and $%
U^{-1}|\tau \rangle $. We will use the  notations in \cite{Grover98}, 
$\langle \tau |U | \gamma\rangle=U_{\tau\gamma}$,
$\langle \gamma|V|\tau\rangle=V_{\gamma \tau}$,
$\langle \tau |V^{-1} | \gamma\rangle=V^*_{\gamma\tau}$,
$\langle \tau |(UV) | \tau\rangle=(UV)_{\tau\tau}$,
 $\langle \gamma|(VU) | \gamma\rangle=(VU)_{\gamma\gamma}$.  
 After calculating $Q|\gamma \rangle $,$Q((VU)|\gamma
\rangle )$, $Q(V|\tau \rangle )$ and $Q(U^{-1}|\tau \rangle )$ we obtain the
following express. 
$Q\left( 
\begin{array}{l}
|\gamma \rangle \\ 
(VU)|\gamma \rangle \\ 
V|\tau \rangle \\ 
U^{-1}|\tau \rangle%
\end{array}%
\right) =$$M$
$ \left( 
\begin{array}{l}
|\gamma \rangle \\ 
(VU)|\gamma \rangle \\ 
V|\tau \rangle \\ 
U^{-1}|\tau \rangle%
\end{array}%
\right) ,$ where

\noindent $M=$
 $\left( 
\begin{array}{llll}
2\cos \theta e^{i\theta }((VU)_{\gamma \gamma }-2\cos \phi e^{i\phi }U_{\tau \gamma }V_{\gamma \tau })
 & -1 & 2\cos \phi e^{i\phi }U_{\tau \gamma } & 0
\\ 
-1+2\cos \theta e^{i\theta }-4\cos \theta e^{i\theta }\cos \phi e^{i\phi
}|V_{\gamma \tau }|^{2} & 0 & 2\cos \phi e^{i\phi }V_{\gamma \tau }^{\ast }
& 0 \\2\cos \theta e^{i\theta }U_{\tau \gamma }^{\ast }-4\cos \theta e^{i\theta }\cos \phi e^{i\phi }V_{\gamma \tau }(UV)_{\tau \tau } 
& 0 & 2\cos \phi e^{i\phi }(UV)_{\tau \tau } 
& -1 \\ 2\cos \theta e^{i\theta }(1-2\cos \phi e^{i\phi })V_{\gamma \tau } & 0 & 
2\cos \phi e^{i\phi }-1 & 0%
\end{array}%
\right)$.

\subsection{The iterated formula for the amplitude}

Next we derive the iterated formula for the amplitudes in the states $|\gamma
\rangle $, $(VU)|\gamma \rangle $,$V|\tau \rangle $ and $U^{-1}|\tau \rangle 
$ for $Q^k |\gamma\rangle $ after $k$ operations of $Q$.

Let $Q|\gamma \rangle =l_{1}|\gamma \rangle -(VU)|\gamma \rangle +p_{1}$$%
(V|\tau \rangle )$,

$\ \ \ Q((VU)|\gamma \rangle )=l_{2}|\gamma \rangle +p_{2}(V|\tau \rangle $,

$\ \ \ Q(V|\tau \rangle )=l_{3}|\gamma \rangle +p_{3}(V|\tau \rangle
)-U^{-1}|\tau \rangle $,

\ \ $\ Q(U^{-1}|\tau \rangle )=l_{4}|\gamma \rangle +p_{4}(V|\tau \rangle ).$

Let $Q^{k}|\gamma \rangle =a_{k}|\gamma \rangle +b_{k}((VU)|\gamma \rangle
)+c_{k}(V|\tau \rangle )+d_{k}(U^{-1}|\tau \rangle )$, where $a_{k}$,$b_{k}$%
, $c_{k}$ and $d_{k}$ are the amplitudes in the states $|\gamma \rangle $, $%
((VU)|\gamma \rangle )$, $(V|\tau \rangle )$ and $(U^{-1}|\tau \rangle )$,
respectively. Then $Q^{k+1}|\gamma \rangle =Q(Q^{k}|\gamma \rangle
)=a_{k}Q|\gamma \rangle +b_{k}Q((VU)|\gamma \rangle )+c_{k}Q(V|\tau \rangle
)+d_{k}Q(U^{-1}|\tau \rangle ).$ Then we obtain the following iterated formula:$%
a_{k+1}=l_{1}a_{k}+l_{2}b_{k}+l_{3}c_{k}+l_{4}d_{k}$ , $b_{k+1}=-a_{k}$ , $%
c_{k+1}=p_{1}a_{k}+p_{2}b_{k}+p_{3}c_{k}+p_{4}d_{k}$ , $d_{k+1}=-c_{k}$ .

\subsection{The first-order approximate formula for the amplitude }

From the iterated formula we will approximate $c_{k+1}$ using the
first-order Taylor formula of $U_{\tau \gamma }$. That is, in $c_{k+1}$ we
only keep the first-order of $U_{\tau \gamma }$ and omit the high order of $%
U_{\tau \gamma }$. Let $V_{\gamma \tau }=U_{\tau \gamma }^{\ast }$, $%
(VU)_{\gamma \gamma }=(UV)_{\tau \tau }=0$ to approximate $c_{k+1}$.
 Let's give a brief justification as follows. Clearly if $V=\pm U^{-1},$ then $%
(VU)=\pm I$, then $(VU)_{\gamma \gamma }=(UV)_{\tau \tau }=\pm 1$. Therefore
assume that $V$ $\neq \pm U^{-1}.$Let $w_{ij}$ be any term of a unitary
matrix $W.$ Then by the definition $WW^{+}=W^{+}W=I\cite{Birkhoff},$ $\sum%
\limits_{k}^{{}}w_{ki}{\bar{w}_{kj}}=\sum\limits_{k}^{{}}w_{ik}{\bar{w}_{jk}}%
=\delta _{ij},$where $\delta _{ij}$ is the Kronecker deta and ${\bar{w}_{kj}}
$ is the complex conjugate of $w_{kj}.$ That is, each row(column) of $W$ has
length one and any two rows(columns) of $W$ are orthogonal. Clearly $%
|w_{ij}|\leq 1$. If $|w_{ij}|=1,$ then all terms of the ith row and the jth
column of $W$ are zero except the $ij$ term. The results above hold for $V
$, $U$, $(VU)$ and $(UV)$ since they are unitary. 
Specially  $|(VU)_{\gamma \gamma }|\leq 1\ $ and $\ |(UV)_{\tau \tau }|\leq 1$.

If $W$ is assumed to be hermitian \cite{Birkhoff},
 then by the definition $W=W^{+},$that is, $w_{ij}={%
\bar{w}_{ji}},$ then all the diagonal terms $ii$ of $W$ are real.
Since $V$ is unitary and $(VU)$
is assumed to be hermitian, it is easy to show that $(UV)$  also is 
hermitian. By the definition $(VU)=(VU)^+=U^+V^+$, then $V^+(VU)V=V^+(U^+V^+)V$,
$UV=V^+U^+=(UV)^+$, therefore $(UV)$ also is hermitian.
So all the diagonal terms of $(VU)$ and $(UV)$ are real.
Specially $(VU)_{\gamma \gamma }$ and $(UV)_{\tau \tau}$ are real.

Next we will give a sufficient condition in which
all the diagonal terms of $(VU)$ and $(UV)$ are zero.
 For the detail please see the appendix 4.

If $V$ is unitary and each $V_{ij}$ of the block form of $V$ is of the form 
$\left( 
\begin{tabular}{ll}
$a$ & $b$ \\ 
$b$ & $a$%
\end{tabular}%
\right) $,
and $U=V^+P$.
 Then $U$ is unitary, $(VU)$ and $(UV)$ are unitary and hermitian, and all the diagonal terms of $(VU)$ and $(UV)$ are zero. For the definition of $P$ please see the appendix 4.

Now let us  make the first-order approximate formula for the amplitude. 
Since $p_{1}$ and $p_{2}$ contain the factor $%
U_{\tau \gamma }$, we have to approximate $a_{k}$ with the expression without $%
U_{\tau \gamma }.$From $Q|\gamma \rangle $ and $Q^{2}|\gamma \rangle $ $%
a_{1}\doteq 0$, $b_{1}=-1$, $c_{1}=2\cos \phi e^{i\phi }U_{\tau \gamma }$, $%
d_{1}=0$; $a_{2}\doteq -e^{i2\theta }$, $b_{2}\doteq 0$, $c_{2}\doteq -2\cos
\phi e^{i\phi }U_{\tau \gamma }$, $d_{2}=-2\cos \phi e^{i\phi }U_{\tau
\gamma }$. By induction we obtain $a_{k+1}\doteq -e^{i2\theta }a_{k-1}$, 
$a_{k}\doteq \cases{0, &$k$ is odd;\cr
                     (-1)^{m}e^{i2m\theta },&$k=2m$.\cr}$ 
 $ c_{k+1}\doteq 2\cos \phi e^{i\phi }(a_{k}-a_{k-1})U_{\tau \gamma }+(1-2\cos
\phi e^{i\phi })c_{k-1}$,
 $c_{k+1}\doteq \cases{(-1)^{m}2\cos \phi
e^{i\phi }U_{\tau \gamma }\sum\limits_{l=0}^{m}\sigma ^{(m-l)}\delta ^{l},&$k
=2m$;\cr
(-1)^{m+1}2\cos \phi e^{i\phi }U_{\tau \gamma
}\sum\limits_{l=0}^{m}\sigma ^{(m-l)}\delta ^{l},&$k=2m+1$,\cr}$ 

where $\sigma =2\cos \theta e^{i\theta }-1=e^{i2\theta },$ $\delta =2\cos \phi e^{i\phi
}-1=e^{i2\phi }$ .

Case 1, $\sigma =\delta .$That is, $\theta =\phi $, which is called phase
matching condition by Long  et al.'s terminology. 

In the case $c_{k+1}\doteq \cases{(-1)^{m}(k+2)\cos \phi e^{i\phi }\sigma ^m U_{\tau \gamma
},&$k=2m$;\cr
(-1)^{m+1}(k+1)\cos \phi e^{i\phi }\sigma ^m U_{\tau \gamma },&$k=2m+1$,\cr}$

\noindent  $|c_{k+1}|\doteq \cases{(k+2)|\cos \phi || U_{\tau \gamma
}|,&$k=2m$;\cr
(k+1)|\cos \phi || U_{\tau \gamma }|,&$k=2m+1$.\cr}$

Therefore for arbitrary unitary
operators $V$ and $U$ , when $(VU)$ is hermitian, $Q=-I_{\gamma }VI_{\tau }U$
\ can be used to construct a quantum search algorithm that succeeds with
certainty except that $I_{\gamma }=I$ or $I_{\tau }=I.$ The number of
iterations is  almost  $1/\left| \cos \phi \right| |U_{\tau \gamma }|$ to
reach the desired state $|\tau \rangle $ from the initial state $|\gamma
\rangle .$

Specially when $\sigma =\delta =1,$that is, $\theta =\phi =0,$ then $%
I_{\gamma }=I_{\gamma }^{(\pi )},I_{\tau }=I_{\tau }^{(\pi )}$, the
algorithm becomes $-I_{\gamma }^{(\pi )}VI_{\tau }^{(\pi )}U,$ and 

\noindent $Q\left( 
\begin{array}{l}
|\gamma \rangle \\ 
(VU)|\gamma \rangle \\ 
V|\tau \rangle \\ 
U^{-1}|\tau \rangle%
\end{array}%
\right) =\left( 
\begin{array}{llll}
(2(VU)_{\gamma \gamma }-4U_{\tau \gamma }V_{\gamma \tau })
 & -1 & 2U_{\tau
\gamma } & 0 \\ 
(1-4|V_{\gamma \tau }|^{2}) & 0 & 2V_{\gamma \tau }^{\ast } & 0 \\ 
(2U_{\tau \gamma }^{\ast }-4V_{\gamma \tau }(UV)_{\tau \tau })
& 0 & 2(UV)_{\tau \tau } & -1 \\ 
-2V_{\gamma \tau } & 0 & 1 & 0%
\end{array}%
\right) \left( 
\begin{array}{l}
|\gamma \rangle \\ 
(VU)|\gamma \rangle \\ 
V|\tau \rangle \\ 
U^{-1}|\tau \rangle%
\end{array}%
\right) $.

We can derive the above independently, please see the appendix 2.

Case 2, $\sigma \neq \delta $, that is, $\theta $ $\neq \phi $. $%
|c_{k+1}|\doteq 2\left| \cos \phi \right| |U_{\tau \gamma }|\left| \frac{%
\sin (m+1)(\theta -\phi )}{\sin (\theta -\phi )}\right| ,$ where $k=2m$ or $%
2m+1.$Clearly $\lim\limits_{\theta \rightarrow \phi }|c_{k+1}|=\cases
{(k+2)|U_{\tau \gamma }||\cos \phi |,&$k=2m$;\cr
 (k+1)|U_{\tau \gamma }||\cos \phi |,&$k=2m+1$.\cr}$
 It means when almost $|\theta -\phi |<2|\cos \phi
||U_{\tau \gamma }|$, $Q$ can be used as a quantum search algorithm that
succeeds with certainty though $\theta $ $\neq \phi $.

\section{The  precise formula
for the amplitude for the general algorithms which preserve a
two-dimensional vector space
 and the  non-symmetric effects of different rotating angles}

\bigskip

Though in the algorithm $Q=-I_{\gamma }VI_{\tau }U$ in the section 1 above
let $V=U^{-1}$ then obtain the algorithm in this section $Q=-I_{\gamma
}U^{-1}I_{\tau }U$, note that in the section 1 when we derived the iterated
and approximate formulas of the amplitudes in the state $V|\tau \rangle $ we
assumed $V_{\gamma \tau }=U_{\tau \gamma }^{\ast }$, $(VU)_{\gamma \gamma
}=(UV)_{\tau \tau }=0.$ And in the section 1 $Q$ preserves the four -
dimensional vector space spanned by $|\gamma \rangle $, $(VU)|\gamma \rangle 
$, $V|\tau \rangle $ and $U^{-1}|\tau \rangle .$In this section we will show
that $Q$ preserves the two-dimensional vector space spanned by $|\gamma
\rangle $ and $U^{-1}|\tau \rangle ,$ clearly $V_{\gamma \tau }$, $%
(VU)_{\gamma \gamma }$ and $(UV)_{\tau \tau }$ don't appear. So the results
in this section can not be obtained from the ones in the section 1 by simply
letting $V=U^{-1}$.\ So for the algorithm in this  section it is necessary to
derive its iterated formula, precise one and approximate one of the amplitude in
the desired state and discuss its amplitude amplification.

Let's study the algorithm $Q=-I_{\gamma }U^{-1}I_{\tau }U$, where $I_{\gamma
}=I-2\cos \theta e^{i\theta }|\gamma \left\rangle |\right\langle \gamma |$
and $I_{\tau }=I-2\cos \phi e^{i\phi }|\tau \left\rangle |\right\langle \tau
|.$ When $\theta =\phi =0,$ it reduces to Grover's algorithm.

After calculating, we obtain the following expression. For detailed derivation,
please see the appendix 3.

$Q\left( 
\begin{array}{l}
|\gamma \rangle \\ 
U^{-1}|\tau \rangle%
\end{array}%
\right) =\left( 
\begin{tabular}{ll}
$\alpha $ & $\beta $ \\ 
$\lambda $ & $\delta $%
\end{tabular}%
\ \right) \left( 
\begin{array}{l}
|\gamma \rangle \\ 
U^{-1}|\tau \rangle%
\end{array}%
\right) $

where $\alpha =-(1-2\cos \theta e^{i\theta }+4\cos \theta e^{i\theta }\cos
\phi e^{i\phi }|U_{\tau \gamma }|^{2}),$ $\beta =2\cos \phi e^{i\phi
}U_{\tau \gamma },$ $\lambda =$ $2\cos \theta e^{i\theta }(1-2\cos \phi
e^{i\phi })U_{\tau \gamma }^{\ast },$ $\delta =2\cos \phi e^{i\phi }-1$.
Clearly the present algorithm $Q$ like Grover's and Long et al.'s algorithms
preserves the vector space spanned by $|\gamma \rangle $ and $U^{-1}|\tau
\rangle .$

For Grover's algorithm $\alpha =1-4|U_{\tau \gamma }|^{2},\beta =2U_{\tau
\gamma }$, $\lambda =-2U_{\tau \gamma }^{\ast }$, $\delta =1$.

For Long et al.'s algorithm $\alpha =-e^{i\theta }-($ $-e^{i\theta
}+1)(-e^{i\phi }+1)|U_{\tau \gamma }|^{2}$, $\beta =(-e^{i\phi }+1)U_{\tau
\gamma }$, $\lambda =($ $-e^{i\theta }+1)e^{i\phi }U_{\tau \gamma }^{\ast }$, 
$\delta =-e^{i\phi }.$

Hoyer's algorithm  preserves a two-dimensional vector space
spanned by another two states, where  $\alpha
=-\{(1-e^{i\phi })a+e^{i\phi }\},\beta =(1-e^{i\phi })\sqrt{a}\sqrt{1-a}%
e^{i\varphi },\lambda =(1-e^{i\phi })\sqrt{a}\sqrt{1-a}$, $\delta
=\{(1-e^{i\phi })a-1\}e^{i\varphi }.$

 The process of  deriving the following iterated formula and 
precise formula for the amplitude is the same 
for the present algorithm and Grover' and Long et al.'s and Hoyer's
algorithms and any other quantum search algorithms which preserve a
two-dimensional vector space, 
this is because the derivation is not concerned
in the contents of $\alpha $, $\beta $, $\lambda $ and $\delta .$

\subsection{The iterated formula for the amplitude}
Let's derive the iterated formula for the amplitude in the desired state $%
U^{-1}|\tau \rangle $ of $Q^{k}|\gamma \rangle $. Let 

$Q|\gamma \rangle
=\alpha |\gamma \rangle +\beta (U^{-1}|\tau \rangle ),$ $Q(U^{-1}|\tau
\rangle )=\lambda |\gamma \rangle +\delta (U^{-1}|\tau \rangle ).$ Then 

$Q^{2}|\gamma \rangle =(\alpha ^{2}+\beta \lambda )|\gamma \rangle +\beta
(\alpha +\delta )(U^{-1}|\tau \rangle ),$......(1)

$Q^{3}|\gamma \rangle =(\alpha ^{3}+\beta \lambda (2\alpha +\delta ))|\gamma
\rangle +\beta (\alpha ^{2}+\alpha \delta +\delta ^{2}+\beta \lambda
)(U^{-1}|\tau \rangle ),$...(2)

$Q^{4}|\gamma \rangle =(\alpha ^{4}+\beta \lambda (3\alpha ^{2}+2\alpha
\delta +\delta ^{2})+(\beta \lambda )^{2})|\gamma \rangle $

$\ \ \ \ \ \ \ +\beta (\alpha ^{3}+\alpha ^{2}\delta +\alpha \delta
^{2}+\delta ^{3}+2(\alpha +\delta )\beta \lambda )(U^{-1}|\tau \rangle )$%
...(3)

Let $Q^{k}|\gamma \rangle =a_{k}|\gamma \rangle +b_{k}(U^{-1}|\tau \rangle )$%
, where $a_{k}$ and $b_{k}$ are the amplitudes in the initial state $|\gamma \rangle 
$ and the desired state $U^{-1}|\tau \rangle $, respectively$.$ Then $%
Q^{k+1}|\gamma \rangle =Q(Q^{k}|\gamma \rangle )=a_{k}Q|\gamma \rangle
+b_{k}Q(U^{-1}|\tau \rangle )$

$=(\alpha a_{k}+\lambda b_{k})|\gamma \rangle +(\beta a_{k}+\delta
b_{k})(U^{-1}|\tau \rangle ),$clearly $a_{k+1}=(\alpha a_{k}+\lambda b_{k})$
and $b_{k+1}=(\beta a_{k}+\delta b_{k}).$

It is also the iterated formula for amplitude amplification  for
Grover's and Long  et al.'s and Hoyer's  algorithms and any other quantum
search algorithm which preserves a two-dimensional vector space.
Clearly it does not need to computer $Q^k(U^{-1}|\tau \rangle )$ to derive 
the iterated formula.

\subsection{The precise formula for the amplitude}
From the iterated formula above by induction  $%
a_{k}$ and $b_{k}$ can be precisely  written
 as the following polynomial  in $(\beta\lambda) $, respectively.
 Let $[x]$ be the greatest integer which is or less than $x.$

$b_{k}=\beta (c_{k0}+c_{k1}(\beta \lambda )+c_{k2}(\beta \lambda
)^{2}+...+c_{k[(k-1)/2]}(\beta \lambda )^{[(k-1)/2]})$, ......(4),

\noindent  where $%
c_{kj}=\sum\limits_{n=k-1-2j}^{0}l_{k(k-1-2j-n)}^{(j)}\alpha ^{n}\delta
^{k-1-2j-n}$,and $l_{ki}^{(j)}=\left( 
\begin{tabular}{c}
i+j \\ 
j%
\end{tabular}%
\right) \left( 
\begin{tabular}{c}
k-i-j-1 \\ 
j%
\end{tabular}%
\right) $.

$a_{k}=\alpha ^{k}+d_{k1}(\beta \lambda )+d_{k2}(\beta \lambda
)^{2}+...+d_{k[k/2]}(\beta \lambda )^{[k/2]}$......(5),

\noindent  where $%
d_{kj}=\sum\limits_{n=k-2j}^{0}t_{k(k-2j-n)}^{(j)}\alpha ^{n}\delta
^{k-2j-n} $, and  $t_{ki}^{(j)}=\left( 
\begin{tabular}{c}
i+j-1 \\ 
j-1%
\end{tabular}%
\right) \left( 
\begin{tabular}{c}
k-i-j \\ 
j%
\end{tabular}%
\right) $.Note that $\left( 
\begin{tabular}{c}
n \\ 
0%
\end{tabular}%
\right) =1,$for any $n\geq 0.$

For Grover's algorithm $(\beta \lambda )=\alpha -1$, $\delta
 =1$. Then $b_k$ can be written
as $b_k=\beta r_k$, where $r_k$ is real.
 
Here it does not need to computer $Q^k(U^{-1}|\tau \rangle )$ to derive the precise formula above. Clearly it will save much more time to computer amplitude amplification in the desired state using the precise formula for $b_k$ than the kth power of the matrix which represents the operator $Q$.
The formula is also the precise formula for the amplitude amplification  for
Grover's and Long  et al.'s and Hoyer's  algorithms and any other quantum
search algorithm which preserves a two-dimensional vector space.
From the precise formula for $b_{k}$, it is not hard to see the
phase conditions in \cite{Hoyer} are only sufficient.

For the detailed derivation of $b_{k}$, please see appendix 3.

\subsection{ The  non-symmetric effects of different rotating angles}
Let us study  the  non-symmetric effects of different rotating angles
of the initial state $|\gamma \rangle $ and the desired state $U^{-1}|\tau
\rangle $ on the amplitude in the desired state.
Clearly the norm of $b_{k}$ contains 
the factor $|\beta |$ which is $2|\cos \phi |$ $|U_{\tau
\gamma }|.$It is not hard to see that the effect of $\phi $ on the amplitude
in the desired state is greater than that of $\theta $ when $\theta \neq \phi
.$It means that when $\theta \neq \phi $ the effects of $\theta $ and $\phi $
on the amplitude in the desired state are not symmetric. 
 For example, when $\phi =\pi /2$, then $I_{\tau
}=I$,$Q=-I_{\gamma }U^{-1}I_{\tau }U=-I_{\gamma }U^{-1}U=$ $-I_{\gamma }$,$%
b_{k}=0$, it means that $Q$ does nothing in the amplitude in the desired state except that
it rotates the phase of the initial state $|\gamma \rangle $ by angle $\pi
+2\theta $. When $\theta =\pi /2,$ then $I_{\gamma }=I$,$Q=-I_{\gamma
}U^{-1}I_{\tau }U=-U^{-1}I_{\tau }U$,$Q^{k}|\gamma \rangle =(-1)^{k}|\gamma
\rangle +(\beta \sum\limits_{i=0}^{k-1}(-1)^{k-1-i}\delta ^{i})U^{-1}|\tau
\rangle $, $|b_{k}|=2|U_{\tau \gamma }||\sin k(\pi /2-\phi )|.$
The result in this subsection is also true for Long et al.'s algorithm.

\subsection{The first-order approximate formula for the amplitude }
Long  et al. used many transformations and approximate operations to
derive the approximate formula for the amplitude amplification[4]. Next let's
derive only  by induction the first-order approximate formula for the
amplitude $b_{k}$ in the desired state $U^{-1}|\tau \rangle $ using the
iterated formula above though it is easy to derive it from the polynomial
in $(\beta \lambda) $ of $a_{k}$ and $b_{k}$, please see (4) and (5)above. 
We only keep the first order of $U_{\tau \gamma }$ in the amplitude in the
state $U^{-1}|\tau \rangle $. In $b_{k+1},$ $\beta $ contains the factor $%
U_{\tau \gamma },$ so $a_{k}$ should be approximated with the express
without $U_{\tau \gamma }$; $\delta $ does not contain the term $U_{\tau
\gamma }$, so $\delta b_{k}$ only contains the first order of $U_{\tau
\gamma }$ provided that $b_{k}$ only contains the first order of $U_{\tau
\gamma }$. In $a_{k+1},$ since $\lambda $ contains the factor $U_{\tau
\gamma }^{\ast }$ and $b_{k}$ contains the factor $U_{\tau \gamma },$ $%
\lambda b_{k}$ must contain $|U_{\tau \gamma }|^{2}$ and is omitted.
Therefore $a_{k+1}$ should be approximated by $\sigma $ $a_{k}$, that is, $%
a_{k+1}\doteq \sigma $ $a_{k},$ to make $a_{k+1}$ not contain factor $%
U_{\tau \gamma }.$  Let's see
how to approximate $Q^{k}|\gamma \rangle $ by induction. Clearly

$Q|\gamma \rangle \doteq \sigma |\gamma \rangle +\beta (U^{-1}|\tau \rangle
),$

$Q^{2}|\gamma \rangle \doteq \sigma ^{2}|\gamma \rangle +\beta (\sigma
+\delta )(U^{-1}|\tau \rangle ).$

Assume that $Q^{k}|\gamma \rangle \doteq \sigma ^{k}|\gamma \rangle +\beta
(\sum\limits_{i=0}^{k-1}\sigma ^{k-1-i}\delta ^{i})(U^{-1}|\tau \rangle ).$
Then $Q^{k+1}|\gamma \rangle \doteq \sigma ^{k+1}|\gamma \rangle +(\beta
\sigma ^{k}+\delta \beta \sum\limits_{i=0}^{k-1}\sigma ^{k-1-i}\delta
^{i})(U^{-1}|\tau \rangle )=\sigma ^{k+1}|\gamma \rangle +\beta
(\sum\limits_{i=0}^{k}\sigma ^{k-i}\delta ^{i})(U^{-1}|\tau \rangle ).$ Then
by induction $b_{k}\doteq $\  $\beta \sum\limits_{i=0}^{k-1}\sigma
^{k-1-i}\delta ^{i}$. 
It is easy to verify that the approximate formula is
just the first-order  of $U_{\tau \gamma }$ 
in  the precise formula for  $b_k$. Please see the appendix 3.
 Clearly it does not need to computer 
$Q^k(U^{-1}|\tau \rangle )$ to derive the approximate formula above.

\subsection{ A mathematical proof
of Grover's algorithm being optimal 
among the  algorithms with arbitrary phase rotations}
Next let's use the approximate formula to study its amplitude amplification
and prove that Grover's algorithm is optimal. Clearly $|\beta
\sum\limits_{i=0}^{k-1}\sigma ^{k-1-i}\delta ^{i}|\leq 2k$ $|U_{\tau \gamma
}|.$We will prove that for any case the norm of amplitude in desired state $%
U^{-1}|\tau \rangle $ after $k$ operations of $Q$ is less than $2k$ $%
|U_{\tau \gamma }|$ except Grover's algorithm.

Case 1, $\sigma =\delta $, that is, $\theta =\phi .$ In the case the
amplitude $b_{k}\doteq $ $2k\cos \phi e^{i\phi }\sigma ^{k-1}U_{\tau \gamma }
$, $|b_{k}|\doteq 2k|\cos \phi |$ $|U_{\tau \gamma }|$. When $\cos \phi \neq
0,$that is, $I_{\gamma }=I_{\tau} \neq I,$we obtain a quantum search
algorithm that succeeds with certainty.

\ \ Case 1.1. When $\sigma =\delta =1,$ that is, $\theta =$ $\phi =0,$that
is Grover's algorithm, $b_{k}\doteq $ $\beta \sum\limits_{i=0}^{k-1}\sigma
^{k-1-i}\delta ^{i}=2kU_{\tau \gamma }$, (please see the appendix 1 to check
it), the norm of the amplitude $|b_{k}|\doteq 2k$ $|U_{\tau \gamma }|$. Let $%
2k$ $|U_{\tau \gamma }|=1,$ we obtained the maximum probability for the
desired state $U^{-1}|\tau \rangle ,$ in the case the optimal number of
operations of $Q$ is $1/2|U_{\tau \gamma }|$ , when $|U_{\tau \gamma }|$
is taken as $1/\sqrt{N},$ the number is $\sqrt{N}/2.$ Here the optimal
number is less than Grover's optimal number $\pi /4|U_{\tau \gamma }|$
evaluated in \cite{Grover98}. The optimal number of iteration steps
obtained by Long et al. \cite{Long1}\ is $\pi
/4\beta ,$where $|U_{\tau \gamma }|=\sin \beta ,$so the optimal number
also is almost $\pi /4|U_{\tau \gamma }|.$ Please see the $|b_{k}|$ in the table 1.

Table 1.

\begin{tabular}{lllll}
$N$ & $\sqrt{N}/2$ & $U_{\tau \gamma }=1/\sqrt{N}$ & $k$ & $|b_{k}|$ \\ 
100 & 5 & 0.1 & 6 & 0.9375 \\ 
400 & 10 & 0.05 & $12$ & 0.9334 \\ 
625 & 12 & 0.04 & $14$ & 0.9010 \\ 
900 & 15 & $1/30$ & $17$ & 0.9064%
\end{tabular}

\ \ Case 1.2.\ When $\sigma =\delta \neq 1,$clearly $|\beta
\sum\limits_{i=0}^{k-1}\sigma ^{k-1-i}\delta ^{i}|<2k$ $|U_{\tau \gamma }|.$

Case 2, $\sigma \neq \delta $, that is, $\theta $ $\neq \phi $. Clearly $%
\sum\limits_{i=0}^{k-1}\sigma ^{k-1-i}\delta ^{i}=\frac{\sigma ^{k}-\delta
^{k}}{\sigma -\delta }$. $|b_{k}|\doteq $ $|\beta
\sum\limits_{i=0}^{k-1}\sigma ^{k-1-i}\delta ^{i}|=2|\cos \phi ||U_{\tau
\gamma }|\sqrt{\frac{1-\cos 2k(\theta -\phi )}{1-\cos 2(\theta -\phi )}=}%
2|\cos \phi ||U_{\tau \gamma }||\frac{\sin k(\theta -\phi )}{\sin (\theta
-\phi )}|$. In the case $\sigma \neq \delta ,$ $\cos (\theta -\phi )\neq \pm
1,$by the induction on $k,$ it is not hard to prove that $|\frac{\sin
k(\theta -\phi )}{\sin (\theta -\phi )}|<k(k>1),$ therefore $|b_{k}|\doteq $ 
$|\beta \sum\limits_{i=0}^{k-1}\sigma ^{k-1-i}\delta ^{i}|<$ $2k$ $|U_{\tau
\gamma }|$.

From the cases 1 and 2, for Grover's algorithm in the first-order
approximate $|b_{k}|=2k$ $|U_{\tau \gamma }|$; for other cases $|b_{k}|<2k$ $%
|U_{\tau \gamma }|$. It proved that Grover's algorithm is an  optimal one with
 the form $-I_{\gamma }U^{-1}I_{\tau }U.$From above $|b_{k}|\doteq 
\cases{2k|\cos \phi ||U_{\tau \gamma }|,&$\theta =\phi$;\cr
       2\left| \cos \phi \right| |U_{\tau \gamma }|\left| \frac{\sin k(\theta -\phi )}{\sin
(\theta -\phi )}\right| ,&$\theta \neq \phi$,\cr}$
 clearly the approximate formula
is simpler than Long et al.'s one \cite{Long1}.

\subsection{The first-order 
approximate phase condition with different rotating angles} 
In \cite{Long2} Long  et al. studied the effects of imperfect phase inversion. In
\cite{Hoyer} Peter Hoyer thinks 
if $\theta $ $\neq \phi $ and $|\varphi -\phi |\leq c /\sqrt{N}$ 
for some approximate constant $c$ then the marked state can still
be found by Long  et al.'s algorithm with high probability. 
 Here we will give the first-order
approximate phase condition that $|\theta -\phi |<2|\cos \phi ||U_{\tau \gamma }|$ and
deduce when $\theta $ $\neq \phi $ and $\theta $ and $\phi $ satisfy the
condition then the desired state can still be found by the present algorithm.

Let $Max_{k}\left| b_{k}\right| $ be the maximal $|b_{k}|$ for any $k$. When 
$\theta $ $\neq \phi $ and $|\theta -\phi |$ is small,$\ Max_{k}\left|
b_{k}\right| \doteq 2|\cos \phi ||U_{\tau \gamma }|/|\theta -\phi |$. It
means that $Max_{k}\left| b_{k}\right| $ is the inverse ratio of $|\theta
-\phi |$ when $|\theta -\phi |$ is small. When $\cos \phi \neq 0,$let $%
|\theta -\phi |=2l|\cos \phi ||U_{\tau \gamma }|$. Then $Max_{k}\left|
b_{k}\right| \doteq 1/l$. When $0\leq l<1,$that is, $%
|\theta -\phi |<2|\cos \phi ||U_{\tau \gamma }|$, almost $Max_{k}\left|
b_{k}\right| \doteq 1;$ when $1<l,$that is, $|\theta -\phi |>2|\cos \phi
||U_{\tau \gamma }|$, $Max_{k}\left| b_{k}\right| \leq 1/l<1.$Please see the
following table 2.In the table 2 let $U_{\tau \gamma }=1/\sqrt{N}$ and $\phi
=0$, where $N=100,U_{\tau \gamma }=0.1$. The experiments were done on IBM PC
using MATLAB.

The table 2. \ \ \ \ \ \ \ \ \ \ \ \ \ \ \ \ \ \ \ \ \ \ \ \ \ \ \ \ \ \ \ \
\ \ \ \ 

\begin{tabular}{lll}
$\theta $ & $k$ & $|b_{k}|$ \\ 
$0.01$ & $7$ & 0.9899 \\ 
0.02 & 8 & 0.9994 \\ 
0.03 & 8 & 0.9930 \\ 
0.04 & 100 & 0.9861 \\ 
0.05 & 100 & 0.9525%
\end{tabular}

When $\sigma \neq \delta ,$ clearly $\lim\limits_{\theta \rightarrow \phi
}2|\cos \phi ||U_{\tau \gamma }||\frac{\sin k(\theta -\phi )}{\sin (\theta
-\phi )}|=$ $2k|\cos \phi ||U_{\tau \gamma }|.$When $\phi =0$, the
limitation is $2k$ $|U_{\tau \gamma }|$, which is just Grover's algorithm.

\bigskip
 {\bf  Acknowledgements}

Thank Prof. G.L. Long for his reading the paper and  discussion with him and comments.
Thank the reviewer for the helpful comments.

{\bf Appendix 1}

For Grover's algorithm,$Q=-I_{\gamma }^{(\pi )}U^{-1}I_{\tau }^{(\pi )}U,$

$Q|\gamma \rangle =(1-4|U_{\tau \gamma }|^{2})|\gamma \rangle +2U_{\tau
\gamma }U^{-1}|\tau \rangle $

$Q^{2}|\gamma \rangle =Q(Q|\gamma \rangle )=((1-4|U_{\tau \gamma
}|^{2})^{2}-4|U_{\tau \gamma }|^{2})|\gamma \rangle +(4U_{\tau \gamma
}-8U_{\tau \gamma }|U_{\tau \gamma }|^{2})(U^{-1}|\tau \rangle ).$

$Q^{3}|\gamma \rangle =Q(Q^{2}|\gamma \rangle )$

$=((1-4|U_{\tau \gamma }|^{2})^{3}-12|U_{\tau \gamma }|^{2}+32|U_{\tau
\gamma }|^{4})|\gamma \rangle +(6U_{\tau \gamma }-32U_{\tau \gamma }|U_{\tau
\gamma }|^{2}+32U_{\tau \gamma }|U_{\tau \gamma }|^{4})(U^{-1}|_{\tau
}\rangle )$

\bigskip

\bigskip

{\bf Appendix 2}

\bigskip

$Q=-I_{\gamma }^{(\pi )}VI_{\tau }^{(\pi )}U$. And note that $\langle \gamma
|\gamma \rangle =1$, $\langle \tau |U|\gamma \rangle =U_{\tau \gamma
},\langle \gamma |U^{-1}|\tau \rangle =U_{\tau \gamma }^{\ast }$ ( Since $%
U^{-1}=U^{\ast }$ and $\langle \gamma |U^{\ast }|\tau \rangle =\langle \tau
|U|\gamma \rangle ^{\ast }=U_{\tau \gamma }^{\ast }).$ Then

$Q|\gamma \rangle =-I_{\gamma }VI_{\tau }U|\gamma \rangle =-(I-2|\gamma
\left\rangle {}\right\langle \gamma |)V(I-2|\tau \left\rangle
{}\right\langle \tau |)U|\gamma \rangle $

$=-(I-2|\gamma \left\rangle {}\right\langle \gamma |)VU|\gamma \rangle
+2(I-2|\gamma \left\rangle {}\right\langle \gamma |)V|\tau \left\rangle
{}\right\langle \tau |U|\gamma \rangle $

$=-(VU)|\gamma \rangle +2|\gamma \left\rangle {}\right\langle \gamma
|VU|\gamma \rangle +2U_{\tau \gamma }(V|\tau \rangle )-4U_{\tau \gamma
}|\gamma \left\rangle {}\right\langle \gamma |V|\tau \rangle $

$=-(VU)|\gamma \rangle +(2(VU)_{\gamma \gamma }-4U_{\tau \gamma }V_{\gamma
\tau })|\gamma \rangle +2U_{\tau \gamma }(V|\tau \rangle )$.

$Q(U^{-1}|\tau \rangle )=-I_{\gamma }VI_{\tau }U(U^{-1}|\tau \rangle
)=-I_{\gamma }VI_{\tau }|\tau \rangle =I_{\gamma }V|\tau \rangle$

$=(I-2|\gamma \left\rangle {}\right\langle \gamma |)V|\tau \rangle =V|\tau
\rangle -2V_{\gamma \tau }|\gamma \rangle .$

$Q(VU|\gamma \rangle )=-I_{\gamma }VI_{\tau }U(VU|\gamma \rangle
)=-(I-2|\gamma \left\rangle {}\right\langle \gamma |)V(I-2|\tau \left\rangle
{}\right\langle \tau |)U(VU|\gamma \rangle )$

$=-(I-2|\gamma \left\rangle {}\right\langle \gamma |)((VUVU)|\gamma \rangle
-2V|\tau \left\rangle {}\right\langle \tau |UVU|\gamma \rangle )$

$=-(VUVU)|\gamma \rangle +2V|\tau \left\rangle {}\right\langle \tau
|UVU|\gamma \rangle +2|\gamma \left\rangle {}\right\langle \gamma
|VUVU|\gamma \rangle -4|\gamma \left\rangle {}\right\langle \gamma |V|\tau
\left\rangle |\right\langle \tau |UVU|\gamma \rangle $

When $VU$ is hermitian, then $VUVU=I$, $VUV=U^{-1}$ and $UVU=V^{-1}.$

Then $Q(VU|\gamma \rangle )=-|\gamma \rangle +2V_{\gamma \tau }^{*}(V|\tau
\rangle )+2|\gamma \rangle -4V_{\gamma \tau }V_{\gamma \tau }^{*}|\gamma
\rangle =(1-4|V_{\gamma \tau }|^{2})|\gamma \rangle +2V_{\gamma \tau
}^{*}(V|\tau \rangle ).$

$Q(V|\tau \rangle )=-I_{\gamma }VI_{\tau }U(V|\tau \rangle )=-(I-2|\gamma
\left\rangle {}\right\langle \gamma |)V(I-2|\tau \left\rangle
{}\right\langle \tau |)U(V|\tau \rangle )$

$=-(I-2|\gamma \left\rangle {}\right\langle \gamma |)((VUV)|\tau \rangle
-2V|\tau \left\rangle {}\right\langle \tau |UV|\tau \rangle )$

$=-\left\{ (VUV)|\tau \rangle -2V|\tau \left\rangle {}\right\langle \tau
|UV|\tau \rangle -2|\gamma \left\rangle {}\right\langle \gamma |VUV|\tau
\rangle +4|\gamma \left\rangle {}\right\langle \gamma |V|\tau \rangle
\langle \tau |UV|\tau \rangle \right\} $

Then $Q(V|\tau \rangle )=-U^{-1}|\tau \rangle +2(UV)_{\tau \tau }(V|\tau \rangle )+(2U_{\tau \gamma }^{\ast }-4V_{\gamma \tau }(UV)_{\tau \tau })|\gamma \rangle $

{\bf Appendix 3}

The algorithm $Q=-I_{\gamma }U^{-1}I_{\tau }U$, where $I_{\gamma }=I-2\cos
\theta e^{i\theta }|\gamma \left\rangle |\right\langle \gamma |$ and $%
I_{\tau }=I-2\cos \phi e^{i\phi }|\tau \left\rangle |\right\langle \tau |.$

$Q|\gamma \rangle =-I_{\gamma }U^{-1}I_{\tau }U|\gamma \rangle =-I_{\gamma
}U^{-1}(I-2\cos \phi e^{i\phi }|\tau \left\rangle |\right\langle \tau
|)U|\gamma \rangle $

 $=-I_{\gamma }(U^{-1}U|\gamma \rangle -2\cos \phi e^{i\phi
}U^{-1}|\tau \left\rangle |\right\langle \tau |U|\gamma \rangle )$

$=-I_{\gamma }(|\gamma \rangle -2\cos \phi e^{i\phi }U_{\tau \gamma
}(U^{-1}|\tau \rangle ))$

$=-((I-2\cos \theta e^{i\theta }|\gamma \left\rangle |\right\langle \gamma
|)|\gamma \rangle -2\cos \phi e^{i\phi }U_{\tau \gamma }(I-2\cos \theta
e^{i\theta }|\gamma \left\rangle |\right\langle \gamma |)(U^{-1}|\tau
\rangle ))$

$=-(1-2\cos \theta e^{i\theta }+4\cos \theta e^{i\theta }\cos \phi e^{i\phi
}|U_{\tau \gamma }|^{2})|\gamma \rangle +2\cos \phi e^{i\phi }U_{\tau \gamma
}(U^{-1}|\tau \rangle ).$

$Q(U^{-1}|\tau \rangle )=-I_{\gamma }U^{-1}I_{\tau }U(U^{-1}|\tau \rangle
)=-I_{\gamma }U^{-1}I_{\tau }|\tau \rangle$

$ =-I_{\gamma }U^{-1}(I-2\cos \phi
e^{i\phi }|\tau \left\rangle |\right\langle \tau |)|\tau \rangle $

$=-I_{\gamma }(U^{-1}|\tau \rangle -2\cos \phi e^{i\phi }U^{-1}|\tau
\left\rangle |\right\langle \tau |\tau \rangle )$

$=-I_{\gamma }(1-2\cos \phi
e^{i\phi })(U^{-1}|\tau \rangle )$

$=-(1-2\cos \phi e^{i\phi })(I-2\cos \theta e^{i\theta }|\gamma \left\rangle
|\right\langle \gamma |)(U^{-1}|\tau \rangle )$

$=-(1-2\cos \phi e^{i\phi
})(U^{-1}|\tau \rangle -2\cos \theta e^{i\theta }|\gamma \left\rangle
|\right\langle \gamma |U^{-1}|\tau \rangle )$

$=-(1-2\cos \phi e^{i\phi })(U^{-1}|\tau \rangle -2\cos \theta e^{i\theta
}U_{\tau \gamma }^{*}|\gamma \rangle )$

$=(2\cos \phi e^{i\phi }-1)(U^{-1}|\tau \rangle )+2\cos \theta e^{i\theta
}(1-2\cos \phi e^{i\phi })U_{\tau \gamma }^{\ast }|\gamma \rangle $

The following is the detailed derivation of the precise formula for the
amplitude amplification.

Let $Q^{k}|\gamma \rangle =a_{k}|\gamma \rangle +b_{k}(U^{-1}|\tau \rangle )$%
, where $a_{k}$ and $b_{k}$ are the amplitudes in the state $|\gamma \rangle 
$ and the desired state $U^{-1}|\tau \rangle $, respectively$.$ Then

$b_{5}=\beta ((\alpha ^{4}+\alpha ^{3}\delta +\alpha ^{2}\delta ^{2}+\alpha
\delta ^{3}+\delta ^{4})+(3\alpha ^{2}+4\alpha \delta +3\delta ^{2})\beta
\lambda +(\beta \lambda )^{2})$

$a_{5}=\alpha ^{5}+(4\alpha ^{3}+3\alpha ^{2}\delta +2\alpha \delta
^{2}+\delta ^{3})\beta \lambda +(3\alpha +2\delta )(\beta \lambda )^{2}$

$b_{6}=\beta ((\alpha ^{5}+\alpha ^{4}\delta +\alpha ^{3}\delta ^{2}+\alpha
^{2}\delta ^{3}+\alpha \delta ^{4}+\delta ^{5})$
\noindent $+(4\alpha ^{3}+6\alpha
^{2}\delta +6\alpha \delta ^{2}+4\delta ^{3})\beta \lambda +3(\alpha +\delta
)(\beta \lambda )^{2})$

$\bigskip a_{6}=\alpha ^{6}+(5\alpha ^{4}+4\alpha ^{3}\delta +3\alpha
^{2}\delta ^{2}+2\alpha \delta ^{3}+\delta ^{4})\beta \lambda$ 
\noindent $+(6\alpha ^{2}+6\alpha \delta +3\delta ^{2})(\beta \lambda )^{2}+(\beta \lambda )^{3}$

From the iterated formula it is not hard to by induction show that $a_{k}$
and $b_{k}$ can be written as the following polynomial 
in $(\beta\lambda) $, respectively.

$b_{k}=\beta (c_{k0}+c_{k1}(\beta \lambda )+c_{k2}(\beta \lambda
)^{2}+...+c_{k[(k-1)/2]}(\beta \lambda )^{[(k-1)/2]})$, ......, 

where $%
c_{kj}=\sum\limits_{n=k-1-2j}^{0}l_{k(k-1-2j-n)}^{(j)}\alpha ^{n}\delta
^{k-1-2j-n}$.

\bigskip $a_{k}=\alpha ^{k}+d_{k1}(\beta \lambda )+d_{k2}(\beta \lambda
)^{2}+...+d_{k[k/2]}(\beta \lambda )^{[k/2]}$......, 

where $%
d_{kj}=\sum\limits_{n=k-2j}^{0}t_{k(k-2j-n)}^{(j)}\alpha ^{n}\delta
^{k-2j-n} $. 

$l_{ki}^{j}$ in the coefficients of $\beta \lambda $ in $%
b_{3},b_{4},b_{5},b_{6},b_{7}$ constitute the following pyramid.

The table 3.

\begin{tabular}{ccccccccccccccccccc}
&  &  &  &  &  &  &  &  & 1 &  &  &  &  &  &  &  & $b_{3}$ &  \\ 
&  &  &  &  &  &  &  & 2 &  & 2 &  &  &  &  &  &  & $b_{4}$ &  \\ 
&  &  &  &  &  &  & 3 &  & 4 &  & 3 &  &  &  &  &  & $b_{5}$ &  \\ 
&  &  &  &  &  & 4 &  & 6 &  & 6 &  & 4 &  &  &  &  & $b_{6}$ &  \\ 
&  &  &  &  & 5 &  & 8 &  & 9 &  & 8 &  & 5 &  &  &  & $b_{7}$ & 
\end{tabular}

Note that the diagonal elements 1,2,3,4,5,..., can be represented by $%
(_{1}^{1}),(_{1}^{2}),(_{1}^{3}),(_{1}^{4}),(_{1}^{5})...$;  
the diagonal elements 1,2,3,4,... times 2 are 2,4,6,8,..., , respectively,...; then
they times 3 are 3,6,9,..., respectively;... 

For example, $l_{70}^{(1)}$, $l_{71}^{(1)}$, $l_{72}^{(1)}$, $l_{73}^{(1)}$  and  $%
l_{74}^{(1)}$  
 in the coefficient $c_{71}$ of $\beta \lambda $ in 
$b_{7}$ can  be represented by $(_{1}^{1})(_{1}^{5})$, $(_{1}^{2})(_{1}^{4})$,
 $(_{1}^{3})(_{1}^{3})$, $%
(_{1}^{4})(_{1}^{2})$ and  $(_{1}^{5})(_{1}^{1})$ respectively.

$l_{ki}^{j}$ in the coefficients of $(\beta \lambda )^{2}$ in $%
b_{5},b_{6},b_{7},b_{8}$ constitute the following pyramid.

The table 4.

\begin{tabular}{ccccccccccccccccccc}
&  &  &  &  &  &  &  &  & 1 &  &  &  &  &  &  &  & $b_{5}$ &  \\ 
&  &  &  &  &  &  &  & 3 &  & 3 &  &  &  &  &  &  & $b_{6}$ &  \\ 
&  &  &  &  &  &  & 6 &  & 9 &  & 6 &  &  &  &  &  & $b_{7}$ &  \\ 
&  &  &  &  &  & 10 &  & 18 &  & 18 &  & 10 &  &  &  &  & $b_{8}$ & 
\end{tabular}

Note that  the diagonal elements 1,3,6,10,..., can be represented by $%
(_{2}^{2}),(_{2}^{3}),(_{2}^{4}),(_{2}^{5}),$... ; then the diagonal elements 
1,3,6,... times 3 are 3,9,18,..., respectively; 
then they times 6 are 6,18,..., respectively;...

For example, $l_{70}^{(2)}$, $l_{71}^{(2)}$ and  $l_{72}^{(2)}$ 
in the coefficient $c_{72}$ of $(\beta \lambda)^{2}$ in $b_{7}$ 
can  be also  represented by
 $(_{2}^{2})(_{2}^{4})$, $(_{2}^{3})(_{2}^{3})$  and $(_{2}^{4})(_{2}^{2})$
respectively.

$l_{ki}^{j}$ in the coefficients of $(\beta \lambda )^{3}$ in $%
b_{7},b_{8},b_{9},b_{10}$ constitute the following pyramid.

The table 5.

\begin{tabular}{ccccccccccccccccccc}
&  &  &  &  &  &  &  &  & 1 &  &  &  &  &  &  &  & $b_{7}$ &  \\ 
&  &  &  &  &  &  &  & 4 &  & 4 &  &  &  &  &  &  & $b_{8}$ &  \\ 
&  &  &  &  &  &  & 10 &  & 16 &  & 10 &  &  &  &  &  & $b_{9}$ &  \\ 
&  &  &  &  &  & 20 &  & 40 &  & 40 &  & 20 &  &  &  &  & $b_{10}$ & 
\end{tabular}

Note that the diagonal elements 1,4,10,20,... can be represented by$%
(_{3}^{3}),(_{3}^{4}),(_{3}^{5}),(_{3}^{6}),$...; the  diagonal elements
1,4,10,... times 4 are 4,16,40,..., respectively; then 
they  times 10 are 10, 40,..., respectively;...

For example, $l_{100}^{(3)}$, $l_{101}^{(3)}$, $l_{102}^{(3)}$ and $l_{103}^{(3)}$ 
in the coefficient $c_{103}$ of $(\beta \lambda
)^{3}$ in $b_{10}$ can  be also represented by 
$(_{3}^{3})(_{3}^{6})$, $(_{3}^{4})(_{3}^{5})$, $(_{3}^{5})(_{3}^{4})$  and  
$(_{3}^{6})(_{3}^{3})$ respectively.

Generally $l_{ki}^{j}$ in the coefficients of $(\beta \lambda )^{k}$ in $%
b_{2k+1},b_{2k+2},b_{2k+3},b_{2k+4},b_{2k+5}$ constitute the following
pyramid.

The table 6.

\noindent
{\tiny
\begin{tabular}{cccccccccccc}
&  &  &  & $(_{k}^{k})(_{k}^{k})$ &  &  &  &  &  &   $b_{2k+1}$ \\ 
&  &  & $(_{k}^{k})(_{k}^{k+1})$ &  & $(_{k}^{k+1})(_{k}^{k})$ &  &  &  &  & 
 $b_{2k+2}$ \\ 
&  & $(_{k}^{k})(_{k}^{k+2})$ &  & $(_{k}^{k+1})(_{k}^{k+1})$ &  & $%
(_{k}^{k+2})(_{k}^{k})$ &  &  &  &   $b_{2k+3}$ \\ 
& $(_{k}^{k})(_{k}^{k+3})$ &  & $(_{k}^{k+1})(_{k}^{k+2})$ &  & $%
(_{k}^{k+2})(_{k}^{k+1})$ &  & $(_{k}^{k+3})(_{k}^{k})$   &  &  & $b_{2k+4}$
\\ 
$(_{k}^{k})(_{k}^{k+4})$ &  & $(_{k}^{k+1})(_{k}^{k+3})$ &  & $%
(_{k}^{k+2})(_{k}^{k+2})$ &  & $(_{k}^{k+3})(_{k}^{k+1})$ &  & $%
(_{k}^{k+4})(_{k}^{k})$ &  &   $b_{2k+5}$%
\end{tabular}}

Therefore we can conclude and prove by induction that $l_{ki}^{(j)}=\left( 
\begin{tabular}{c}
i+j \\ 
j%
\end{tabular}%
\right) \left( 
\begin{tabular}{c}
k-i-j-1 \\ 
j%
\end{tabular}%
\right) .$So we obtain the precise formula for the amplitude $b_{k}$ in the
desired state after operations of the algorithm $Q.$

{\bf Appendix 4}

The sufficient condition in which $(VU)$ and $(UV)$ are hermitian and all the diagonal 
terms of $(VU)$ and $(UV)$ are zero

Let's define the $N \times N$  matrix $P$. Let $p_{ij}$ be any terms of $P$, where 
$p_{(2i-1)(2i)}=1$, $p_{(2i)(2i-1)}=1$ and $p_{ij}=0$ otherwise. Clearly $P$ is unitary and 
hermitian, $p^2=1$, and all the diagonal terms of $P$ are zero. It is not hard to see that $P$ has the block form which is $diag \{ P_1, P_2, ..., P_m\}$, where $m=N/2$ and $P_i $ is
$\left( 
\begin{tabular}{ll}
0 & 1 \\ 
1 & 0%
\end{tabular}%
\right)$ for $i=1,2,...,m$.

It is easy to verify that $(PV)$ means to interchange lines $(2k-1)$ and $(2k)$ of $V$
and $(VP)$ means to interchange columns $(2k-1)$ and $(2k)$ of $V$, where $k=1,2,...,N/2$.

Given that $V$ is unitary. Then if $U=V^+P$ then $U$ is unitary and $(VU)=V(V^+P)=(VV^+)P=P$.
AS well if $V=U^+P$ then $(UV)=(UU^+)P=P$. From $U=V^+P$, obtain $U^+=PV$. From $V=U^+P$, obtain $V=PVP$. Then we conclude the following sufficient condition.

Sufficient condition (Version 1).
 If $V$ is unitary, $V=PVP$ and $U=V^+P$, then $(VU)$ and $(UV)$ are unitary and hermitian, and 
all the diagonal terms of $(VU)$ and $(UV)$ are zero.

Proof. From the discussion above $(VU)=P$. Since $V=PVP$, $(UV)=(V^+P)(PVP)=V^+P^2VP=(V^+V)P=P$. By the property of $P$, clearly the lemma holds.

Let $V$ has the block form 
$\left( 
\begin{tabular}{llll}
$V_{11}$ & $V_{12}$ & .. & $V_{14}$ \\ 
$V_{21}$ & $V_{22}$ & .. & $V_{24}$ \\ 
.. & .. & .. & .. \\ 
$V_{m1}$ & $V_{m2}$ & .. & $V_{mn}$%
\end{tabular}%
\right) $, where $m=N/2$ and each $V_{ij}$ is 2$\times $2 submatrix. 

It is not hard to see that $V=PVP$ if and only if $V_{ij}$ is of the form 
$\left( 
\begin{tabular}{ll}
$a$ & $b$ \\ 
$b$ & $a$%
\end{tabular}%
\right) $ for $1\leq i,j\leq m$. 
   
From this we conclude the version 2 of the sufficient condition.

The sufficient condition (Version 2)

If $V$ is unitary and each $V_{ij}$ of the block form of $V$ is of the form 
$\left( 
\begin{tabular}{ll}
$a$ & $b$ \\ 
$b$ & $a$%
\end{tabular}%
\right) $,
and $U=V^+P$. Then $(VU)$ and $(UV)$ are unitary and hermitian, and all the diagonal terms of $(VU)$ and $(UV)$ are zero.

\end{document}